\documentclass[aps,pra,preprint,showpacs,groupedaddress]{revtex4}
\usepackage{graphicx} 
\usepackage{amsmath}
\usepackage{amssymb}

\begin{document}

\title{Classical aspects of ultracold atom wavepacket motion \\ through microstructured waveguide bends}
\author{M.W.J.~Bromley}
  \email{bromley@phys.ksu.edu}
\author{B.D.~Esry}
  \email{esry@phys.ksu.edu}
\affiliation{Department of Physics, Kansas State University, Manhattan, KS 66506 USA}

\date{\today}

% \vspace{1cm}

\begin{abstract}

The properties of low-density, low-energy matter wavepackets propagating
through waveguide bends are investigated.   Time-dependent quantum mechanical
calculations using simple harmonic oscillator confining potentials
are performed for a range of parameters close to those accessible by
recent ``atom chip''-based experiments.  We compare classical calculations
based on Ehrenfest's theorem to these results to determine whether 
classical mechanics can predict the amount of transverse excitation
as measured by the transverse heating.  The present results thus
elucidate some of the limits for which matter wave propagation
through microstructures can be reliably considered using classical particle motion.

\end{abstract}


\maketitle

\section{Introduction}

Recent experimental progress in atomic physics includes the success
of magnetic microtraps \cite{hinds99a} --- the so-called ``atom chips'' \cite{brugger00a} ---
where combinations of magnetic fields and wires layed on silica substrates
have made possible atom traps, guides and devices all above a single chip \cite{reichel01a,folman02a}.

The strong magnetic field gradients created by microtraps have already proven
their versatility.
Bose-Einstein condensates (BECs) have been created above microchip surfaces
using evaporation \cite{hansel01a,ott01a,schneider03a,jones03a}
and also by using the surface itself as a hot atom knife
to condense a cloud of atoms \cite{reichel99a,harber03a}.
These achievements are significant since a BEC is the most likely source to
feed atom-optical devices as they provide a large number of coherent atoms. 
In fact, the controlled propagation of BECs through atom chip waveguides
has already been demonstrated \cite{hansel01b,leanhardt02a,fortagh03a},
and some sources of decoherence have been identified \cite{henkel03a,henkel03b,jones03a}.

Non-linear interactions in atom optics can be useful \cite{molmer03a},
but recent theoretical investigations have demonstrated that such effects
can potentially degrade atom optical device performance \cite{stickney03a,zhang03a,schroll03a,chen03a}.
Furthermore, despite the possibility of using multi-moded matter
waves \cite{andersson02a,stickney02a,girardeau02a}, the
low-density regime with single-mode propagation is expected to
make atom optical devices much simpler and easier to operate.
With this in mind, previous theoretical investigations of the low-velocity, low-density
matter wave limit have delineated the conditions for single-moded
wave propagation through microstructures
\cite{jaaskelainen02a,jaaskelainen03c,bromley03b,bortolotti04a}.

In the present paper, we further explore the conditions under which
propagation is single-moded, and the extent to which
classical mechanics can predict the transverse excitation.
Specifically, the time-dependent scattering of a wavepacket
through a circular bend is investigated, neglecting atom-atom
interactions.
The curved waveguide is a fundamental system that
forms the basis for many geometries already experimentally
realised, such as multiple circular bends \cite{muller99a},
storage rings \cite{sauer01a}, spiral guides \cite{luo04a} and
stadium shaped traps \cite{wu04a}.  Note that these experiments all
used atoms with relatively high energies, not ground mode atoms, and thus
the atomic motion was modelled using Monte Carlo ensembles of
classical particles \cite{muller99a,sauer01a,luo04a}.

The time-dependent quantum mechanical calculations explore
the low-energy, tight bend limit, as well as regimes not easily accessible
with our previous time-independent calculations of the circular bend
system \cite{bromley03b}: large radii bends and small de Broglie wavelengths.
Classical particle calculations using Ehrenfest's theorem are performed
alongside such wavepacket calculations to highlight the connections between the
two pictures of atom propagation:
\begin{align}
\label{eqn:ehren}
\frac{d \langle x \rangle}{dt}  &= \frac{\langle p_x \rangle}{m} \; , \\
%\begin{split}
  \frac{d\langle p_x \rangle}{dt} &= - \; \Big\langle \frac{\partial V(x,y,z)}{\partial x} \Big\rangle \nonumber \\ \notag
  &\approx - \; \frac{\partial V(\langle x \rangle,\langle y \rangle,\langle z \rangle)}
  {\partial \langle x \rangle} \; .
%\end{split}
\end{align}
That is, the center-of-mass motion of a wavepacket can be approximated by
a \textit{single} trajectory of a classical particle, which we call the
Ehrenfest trajectory.  While Ehrenfest's theorem is generally exact
for simple harmonic oscillators (SHOs), our bend is only an SHO in the transverse
direction.  So, the question asked here is how well the approximation introduced in
Eq.~(\ref{eqn:ehren}) reproduces the exact Ehrenfest's theorem, and
thus the quantum mechanical results.
We are not, however, testing the use of Monte Carlo ensembles of
classical particles to approximate the quantum observables.
J{\"a}{\"a}skel{\"a}inen and Stenholm \cite{jaaskelainen02c}, for example,
have explored such an approach and found excellent agreement for
the ``transverse cooling'' of a wavepacket exiting an abruptly terminating
transverse potential (when quantum reflections do not play a role).

In our calculations, remarkable agreement is seen between
the ``transverse heating'' for the quantum wavepacket and for its Ehrenfest trajectory,
when wavepacket properties such as dispersion
can be neglected (in accordance with our approximation to Ehrenfest's theorem)
and where large quantum numbers are involved (in accordance with
Bohr's correspondence principle).
It will also be seen that the transverse heating of the Ehrenfest trajectories
and the wavepackets are in disagreement with analytic expressions obtained by
Blanchard and Zozulya for high-velocity atoms and large radii bends \cite{blanchard01a}.
The present results provide guidance as to when matter wave propagation
through microstructures can be reliably considered using classical particle motion.

\section{Details of the Calculations}

Whilst there are a variety of atom chip wire configurations
both proposed and experimentally proven \cite{thywissen99a,folman02a},
the same theoretical ansatz found in \cite{bromley03b} was adopted here.
That is, only multiple wire configurations that do not require
external bias fields applied in the plane of the microchip surface
are considered \cite{luo04a}.  The problems relating to Majorana spin-flips
during propagation in the zero-field region of such geometries
are ignored here, although it is noted that Luo \textit{et.al.} \cite{luo04a}
have proposed a rotating potential scheme to avoid these losses.

\begin{figure}[th]
\includegraphics[width=3.5in]{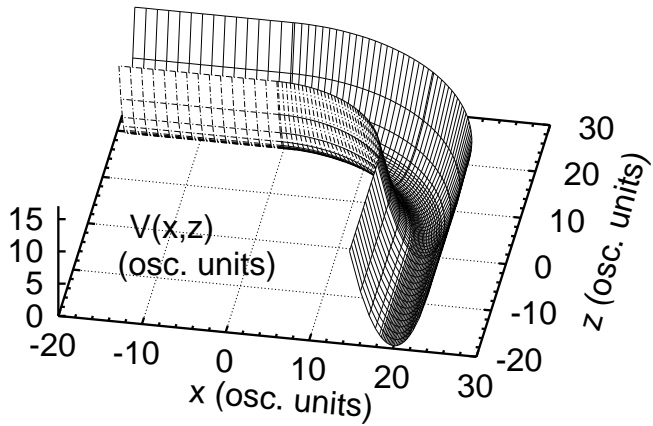}
\caption[]{ \label{fig:circbendpotential}
Potential energy surface for a $\rho_0=20$, $\phi_0=90^\circ$ SHO-based circular
bend.  Both the energy and coordinates are given in oscillator units.
The coordinates are those relative to the $z\le0$ region of Eq.~(\ref{eqn:pots2D}).
}
\end{figure}

Our model reduces exactly to a 2-D geometry without
losing any physics, as the out-of-plane potential remains constant
through the bend in the limit where atom-atom interactions are neglected
(in other words, the out-of-plane quantum number is conserved \cite{bromley03b}).
In practice, guiding potentials are quadratic near their minima,
so a simple harmonic oscillator (SHO) potential is employed here:
\begin{equation}
\label{eqn:pots2D}
V = \begin{cases}
       \frac{1}{2} m \omega^2 (x-\rho_0)^2 & z\le0 \; , \\
       \frac{1}{2} m \omega^2 (\rho-\rho_0)^2 & 0 \le \phi \le \phi_0 \; , \\
       \frac{1}{2} m \omega^2 (x-\rho_0)^2 & z\ge0 \; .
\end{cases}
\end{equation}
Configuration space has been divided into two straight leads described by
Cartesian coordinates and the circular bend connecting them.
The bend has radius $\rho_0$ and angle $\phi_0$, and is described by
polar coordinates.  The potential for a relatively tight ($\rho_0 = 20$),
$90^\circ$ bend is shown in Fig.~\ref{fig:circbendpotential}.

Oscillator units are used throughout this paper (unless indicated otherwise),
where energies are in units of $\hbar\omega$,
lengths are in units of $\beta=\sqrt{\hbar/m\omega}$,
and time is in units of $1/\omega$.
Table \ref{tab:oscunits} shows some typical values for $^{87}$Rb atoms
trapped with a transverse oscillator frequency
of $\omega = 2\pi \times 97$ Hz \cite{leanhardt02a}.
This combination gives $\beta \approx 1.09 \mu$m.   The SHO energy spectrum
and associated velocity, temperature, and wavelength are also shown.
These values correspond to the thresholds of propagation for each mode.
\begin{table}[th]
\caption[]{
\label{tab:oscunits}
Conversion of oscillator units to S.I. units for $^{87}$Rb atoms trapped
by a transverse oscillator frequency $\omega = 2\pi \times 97$ Hz,
\textit{i.e.} an oscillator width $\beta \approx 1.09 \mu$m.
The energy of each SHO mode is given in oscillator units. 
Also given are the velocity, temperature, and wavelength
corresponding to the thresholds of propagation for each mode.
}
\begin{ruledtabular}
\begin{tabular}{cccccc}
   n & $E_{lead}$ & $v_z$ & $E_k$ & $v_z$ & $\lambda$ \\
   & (osc.) & (osc.) & ($\mu K$) & (mm s$^{-1}$) & ($\mu$m)\\
      \hline
   0 & 0.50 & 0           &   0    &    0  &  $\infty$ \\
   1 & 1.50 & $\sqrt{2}$  & 38.71  & 0.943 &  4.86  \\
   2 & 2.50 & $2$         & 77.41  & 1.334 &  3.44  \\
   3 & 3.50 & $\sqrt{6}$  & 116.1  & 1.634 &  2.81  \\
   4 & 4.50 & 2$\sqrt{2}$ & 154.8  & 1.887 &  2.43  \\
  32 & 32.5 & 8           &  1239  & 5.336 &  0.860 \\
\end{tabular}
\end{ruledtabular}
\end{table}

The time-propagation calculations are performed using a
split-operator Crank-Nicolson method with finite differences
on a non-uniform, non-Cartesian grid.  The details of our implementation
of the differencing are discussed in Appendix \ref{sec:finite}
(see also Ref.~\cite{witthoeft03a}),
but we will outline the time propagation scheme below.

\begin{figure}[th]
\includegraphics[width=3.5in]{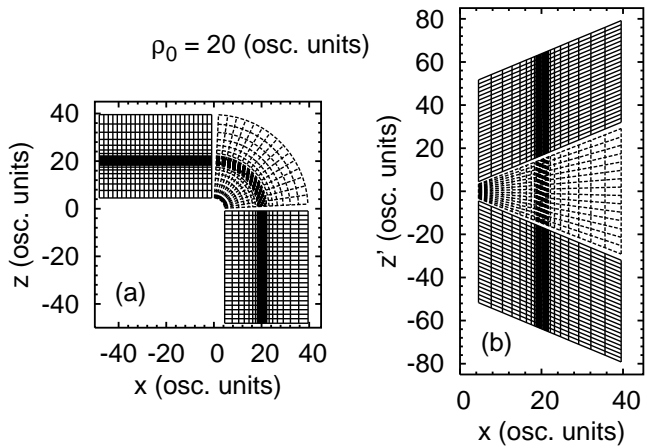}
\caption[]{ \label{fig:circbendgrid}
Example non-Cartesian, non-uniform grid used for the Crank-Nicolson with
finite-differences calculations for a $\phi_0=90^\circ$ circular bend with $\rho_0=20$.
(a) shows the grid points translated to a Cartesian coordinate system.
(b) shows the grid points used in the calculations: where the coordinate
transverse to the direction of propagation in both the leads
and bend is given by $x$, while $z'$ is an auxillary coordinate that measures the
distance from the middle of the bend ($z'=0$ at $\phi=\phi_0/2$) along an
equipotential line.
The grid points located within the bend are connected with the dashed lines for clarity.
}
\end{figure}

We first note that we use a hybrid (Cartesian plus polar)
grid to improve the efficiency and accuracy of the calculations.
A purely Cartesian grid is, in general, computationally wasteful for bends
with angles other than $0^\circ$, so uniform Cartesian grids in
the leads are joined with a uniform polar grid in the bend.  Higher
accuracy is possible by using the coordinate system best adapted to each region.
The same non-uniform transverse grid is used in both the leads
and the bend, with the density of grid points greatest at $\rho_0$, the center of the
potential valley.  An example grid is shown in Fig.~\ref{fig:circbendgrid}
plotted as a function of $z'$, which is an auxillary coordinate that measures
the distance along the guide from the middle of the bend ($z'=0$ at $\phi=\phi_0/2$).
This scheme minimises the number of gridpoints required for
long propagation times through arbitrary angle bends.
In this transformed grid, $2N_z$ plus $N_\phi$ gridpoints are along the direction
of propagation, and the grid spacing $\delta_z$ and $\rho_0\delta_\phi$ are
chosen to be roughly the same.   The significant computational benefit of using
the hybrid grid was that calculations of bends with large radii require essentially the
same number of transverse grid points $N_x$ as smaller bends.
There is a small error introduced where the Cartesian and polar grids are
joined since the differencing there assumes the grid lines are parallel.
This error decreases with increasingly dense grids and with growing $\rho_0$.

The time-dependent solution of the Schr{\"o}dinger equation is
\begin{equation}
\label{eqn:schrosoln}
\Psi(x,z,t+\delta_t) = e^{-iH\delta_t}\Psi(x,z,t) \;.
\end{equation}
One way to evaluate the exponential operator is to write
$H=T+V$ and use the standard split operator idea to write
\begin{equation}
\label{eqn:expandexp}
e^{-iH\delta_t} = e^{-\frac{i}{2}V\delta_t}e^{-iT\delta_t}e^{-\frac{i}{2}V\delta_t} + \mathcal{O}(\delta_t^3) \; .
\end{equation}
The exponentials of the potential are straightforward since the potential
matrix is diagonal, and $e^{-iT\delta_t} = e^{-iT_x\delta_t}e^{-iT_z\delta_t}$ in the
leads since $[T_x,T_z]=0$.  In the bend, $[T_\rho,T_\phi] \ne 0$, and we use 
$e^{-iT\delta_t} \approx e^{-iT_\rho\delta_t/2}e^{-iT_\phi\delta_t} e^{-iT_\rho\delta_t/2}$.

The kinetic energy operators are evaluated as described in Appendix \ref{sec:finite}
and give tridiagonal matrices. Exponentiating these operators is accomplished
with the Crank-Nicolson method.   The time evolution is thus performed in each
spatial direction separately by solving complex tri-diagonal linear equations.
Each timestep requires solving a linear equation $N_x$ times for the $z'$-direction,
and then using the resultant $\Psi$ and solving $2N_z+N_\phi$ linear equations
for the $x$-direction.

The initial wavepacket is a Gaussian centered at $(x,z)=(x_0,z_0)$, with
spatial widths $\Delta_x$ and $\Delta_z$ and initial average velocity $v_z$:
\begin{equation}
\label{eqn:gaussinit}
\Psi(x,z,t=0) = N_0 \; e^{i v_z z} \; e^{-\frac{(x-x_0)^2}{2\Delta_x^2}} \; e^{-\frac{(z-z_0)^2}{2\Delta_z^2}} \; .
\end{equation}
We only consider states that are in the lowest mode of the lead
($x_0=\rho_0$ and $\Delta_x=1$).  Despite a dense transverse grid,
the above analytic ground state is not exactly an eigenstate on the grid, so the normalization constant
$N_0$ is determined numerically from a midpoint integration rule
(see Appendix~\ref{sec:finite}).  The split operator
Crank-Nicolson scheme guarantees the unitarity of the wavefunction for all later times.
Cigar-shaped wavepackets were employed here ($\Delta_z=10$), which
corresponds to the common experimental situation in which a BEC is prepared with
transverse trapping frequencies that are significantly larger than
the frequency in the direction of propagation \cite{leanhardt02a,kasper03a,ott03a}.
They are also used since an elongated Gaussian wavepacket has a
small velocity spread (in oscillator units $\Delta_z\Delta_{v_z}=1$),
reducing the effects of wavepacket dispersion.
The numerical grid is large enough that the
wavepacket propagates through the bend without touching
the edges of the grid during the times of interest.

Expectation values were determined using a midpoint rule
integration (see Appendix \ref{sec:finite}).  In particular,
$\langle H \rangle(t) = \langle T \rangle(t) + \langle V \rangle(t)$ was
monitored to ensure that the total energy variation remained at levels
less than 1 part in $10^8$.  To achieve such accuracy,  a typical
calculation  ($90^{\circ}$ bend with $\rho_0=10$, using a wavepacket
with $v_z=3$, $\Delta_z=10$ starting at $z_0=-200$) required $\delta_t = 0.005$.
We used $N_z = 8000$ covering $-400<z<0$, giving tens of points per wavelength
(determined from the $\lambda=2\pi/v_z$).  In the bend, $N_\phi=314$ points were
chosen to ensure a uniform spacing along the SHO minimum
in both the bend and leads.  The transverse grid ensured that the
oscillations of the maximum SHO mode energetically available (and a few of the closed,
evanescent modes closest to threshold \cite{bromley03b}) could be
accurately described (eg. $N_x=248$ was sufficient for $v_z=3$).

To calculate a classical trajectory through the same bend,
the Runga-Kutta method was employed to solve the classical equations of motion:
\begin{equation}
\label{eqn:classymate}
\frac{d p_x}{dt} = - \frac{\partial V(x,z)}{\partial x} \quad , \quad
\frac{d p_z}{dt} = - \frac{\partial V(x,z)}{\partial z} \; .
\end{equation}

To quantitatively compare the Ehrenfest trajectory and quantum mechanical wavepacket
calculations, the transverse heating $E_h$ is used \cite{blanchard01a}.
The transverse heating measures the amount of propagation energy ($E_z$) transferred into
transverse energy by the bend, eg. for the Ehrenfest trajectory we used
\begin{equation}
\label{eqn:transheat}
E_h = E_z(t\to -\infty) - E_z(t \to \infty) \; .
\end{equation}
If there is no transverse energy initially --- as we assume in the present
calculations --- $E_h$ must be non-negative and there can be no transverse cooling.
Quantum mechanically, we determined the transverse heating of a wavepacket that has been
excited by the bend from the time-averages of $\langle T \rangle(t)$ and $\langle V \rangle(t)$
once the wavepacket had completely exited the bend.  Further
discussion of this is relegated to Appendix \ref{sec:timedeps}.

\section{Results}

In this section we will first discuss a series of $\phi_0=90^\circ$ bends
with various radii $\rho_0$ to highlight the quantum mechanical wavepacket
and Ehrenfest trajectories.  We will then examine the transverse heating
of both wavepackets and Ehrenfest trajectories propagating
through the same bend, and compare these results with the
analytic classical results of Blanchard and Zozulya \cite{blanchard01a}.
The emphasis of the present calculations is on parameters close to currently
realisable experimental situations, \textit{i.e.} larger radii bends and
higher propagation energies.

\subsection{Wavepacket and Ehrenfest trajectories}
\label{sec:wavepacket}

\begin{figure}[th]
\includegraphics[width=3.5in]{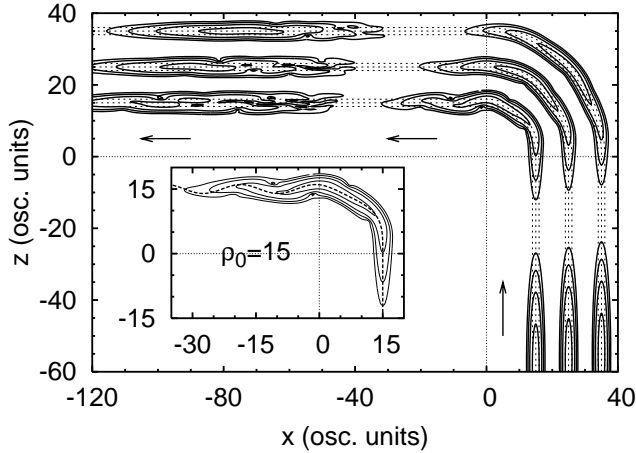}
\caption[]{ \label{fig:projcompilev03000}
Propagation of a wavepacket with $v_z=3$ through three $90^\circ$ circular
bends ($\rho_0=15,25,35$) with the SHO waveguide center marked at $15\pm1$, $25\pm1$ and $35\pm1$.
Three snapshots at roughly the same time are shown, with each contour line
corresponding to a logarithmic decrease of $|\Psi(x,z,t)|^2$
from $10^{-2}$ to $10^{-5}$
(for the initial wavepacket, the peak probability
on the grid was $|\Psi(x,z,t_0)|^2 = 3.18\times10^{-2}$).
No reflections were observed at this level of detail.
Inset: $\rho_0=15$ wavepacket superimposed on its
Ehrenfest trajectory (dashed line).
}
\end{figure}

The calculations in this section are presented to provide an understanding
of the fundamental wavepacket and Ehrenfest trajectory dynamics.  
To demonstrate multimode excitation, an incoming wavepacket with $v_z=3$ was
chosen. At this energy, four excited modes are energetically open.
The excitation of a wavepacket during propagation through
three different $90^\circ$ bends ($\rho_0 = 15,25,35$) can be observed
in Fig.~\ref{fig:projcompilev03000},
which shows the probability density $|\Psi(x,z,t)|^2$ contours of the wavepackets at
three roughly equal times.  In each case, a cigar-shaped wavepacket in the
ground transverse state with $\Delta_z / \Delta_x = 10$ was used, and
was initially located at $z_0=-200$.

The tightest of the three bends yields significant population
of all five available modes in the final snapshot.  Due to energy conservation,
the conversion of longitudinal kinetic energy into transverse energy (heating)
means that each excited mode has a different average propagation velocity.
Consequently, the higher modes lag further and further behind the ground mode
as time progresses.   This effect can be seen in the final snapshots
in Fig.~\ref{fig:projcompilev03000}, where the excited modes are showing
the first signs of separating from the ground state.
It should be emphasized that the number of modes
available is the same for all $\rho_0$ shown in Fig.~\ref{fig:projcompilev03000}.
The degree of mode-excitation, however, depends on $\rho_0$. 
The two main conclusions of our time-independent calculations
for circular waveguide bends \cite{bromley03b} are also bourne out here.
That is, in general, there is minimal reflection from circular bends,
and mode transfer can become significant for tight bends.

\begin{figure}[th]
\includegraphics[width=3.5in]{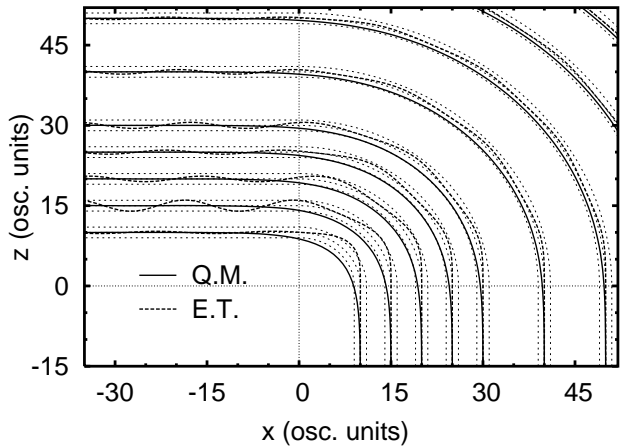}
\caption[]{ \label{fig:bendcircposy}
Ehrenfest trajectories (E.T., dashed lines) through $90^\circ$ bends for
various $\rho_0$, with fixed incoming velocity $v_z=3$ (incoming from the
bottom of the figure).
The solid lines correspond to the path of
$\langle x \rangle(t)$,$\langle z \rangle(t)$
for the time-dependent wavepacket calculations (Q.M.). The dotted lines are
the positions of the SHO minima $\rho_0$ and $\rho_0 \pm 1$.
}
\end{figure}

To determine the suitability of classical mechanics for this problem, we
invoke Ehrenfest's theorem within the approximation shown in Eq.~(\ref{eqn:ehren}).
Given the exact agreement between quantum and classical mechanics Ehrenfest's
theorem usually gives for SHOs, it might well be expected that this approach would
be sufficient for the present problem \cite{cornell03p}.  Ehrenfest's theorem
requires that we solve the classical equations of motion with initial conditions
matching the expectation values of position and velocity for the corresponding
wavepacket.  We thus need only consider a single classical trajectory for
each wavepacket, the Ehrenfest trajectory.
For the present initial wavepacket ($\Delta_x=1$, $\Delta_z=10$, $v_z=3$,
$x_0 = \rho_0$, $z_0=-200$), the trajectory begins at
$x_0 = \rho_0$, $z_0 \le 0$ with $v_z=3$, $v_x=0$ and the
resulting Ehrenfest trajectories for a series of circular $90^\circ$ bends
can be seen in Fig.~\ref{fig:bendcircposy}.  The paths of
the wavepacket average position $\langle x \rangle(t)$,
$\langle z \rangle(t)$ from the time-dependent quantum calculations are also shown.

The Ehrenfest trajectory \textit{always} maintains $\rho(t) \ge \rho_0$ in
the bend, even though the SHO minimum is at $\rho_0$.  This behavior is due to
the conversion of linear momentum into angular momentum $\ell_y=v_z\rho_0$ in
the bend, which results in an effective transverse potential,
\begin{equation}
\label{eqn:veff}
V_{eff}(\rho) = \frac{1}{2}(\rho-\rho_0)^2 + \frac{\ell^2_y}{2\rho^2} \; .
\end{equation}
Since the Ehrenfest trajectory enters with zero transverse kinetic energy,
this potential has an inner turning point at $\rho=\rho_0$
(from $V_{eff} = E_z = v_z^2/2$), and an outer turning point at
$\rho=\rho_0+4E_z/\rho_0+{\cal O}(\rho_0^{-2})$ for large $\rho_0$.  Thus, the Ehrenfest
trajectory always exits a circular bend with $\rho\ge\rho_0$, and
the outer turning point approaches $\rho=\rho_0$ for large $\rho_0$.
In general, this means that the amount of propagation energy transferred
into transverse energy decreases with increasing $\rho_0$.

In the quantum mechanical situation, the effect of the centrifugal barrier
is manifested in the transverse eigenmodes within the bend:
they slosh outwards with increasing energy \cite{bromley03b}.
Nevertheless, the path of a wavepacket's average position is such
that it always lies at $\langle \rho \rangle(t) \le \rho_0$.
This difference from classical mechanics stems from using an elongated initial wavepacket.
For the smallest bends, the wavepacket is comparable to the bend
length ($\Delta_z(t) \sim \rho_0\phi_0$), and the difference is most
exaggerated.  As $\rho_0$ increases, with everything else held fixed,
the Ehrenfest trajectories and the path of the wavepacket's average
position do approach one another at $\rho=\rho_0$ (particularly obvious
in the top-right corner of Fig.~\ref{fig:bendcircposy} in the
$\rho_0=60$ and $70$ calculations).

We can see that Ehrenfest's theorem does not give the agreement
that one might expect for a SHO.  This discrepancy can be understood by
examining Eq.~(\ref{eqn:ehren}).  For a SHO, $\langle -\partial V/\partial x\rangle =
   \langle -m \omega x \rangle = -m\omega\langle x \rangle$,
so Ehrenfest's theorem --- and our approximation to it --- is exact.
A single classical trajectory will thus exactly reproduce the
quantum wavepacket's expectation values.  In the present case,
however, we get no such simple relation because of the bend, and must
approximate $\langle -\partial V(x,z)/\partial x\rangle \approx
-\partial V(\langle x \rangle,\langle z \rangle)/\partial \langle x \rangle$.
This approximation makes Ehrenfest's theorem tractable, but also leads
to its eventual breakdown.  Without this approximation, Ehrenfest's theorem
would be exact for all bend parameters.

Despite the discrepancy between the quantum and classical paths,
a feature of these systems is best illustrated
in the $\rho_0=15$ inset of Fig.~\ref{fig:projcompilev03000}, where
the probability density of the wavepacket is superimposed on
its Ehrenfest trajectory.   This shows that, in the tight bend limit,
the wriggles of a wavepacket exiting the bend do follow its Ehrenfest
trajectory.  The wavepacket wriggles are due
to a superposition of transverse modes (see Appendix \ref{sec:timedeps}) and,
in some sense, the wriggles form the wavepacket's ``trajectory'' as they are
the path of maximum wavefunction flux.
To further compare and contrast a wavepacket and its Ehrenfest trajectory,
we next examine the amount of transverse heating by a series of circular bends.

\subsection{Transverse Heating}
\label{sec:classy}

The transverse heating is an experimental observable.
It has been measured, for example, for a beam of cold atoms propagating
through multiple circular bends \cite{muller99a}.
Theoretically, transverse heating created by curved waveguides was
investigated by Blanchard and Zozulya (BZ) \cite{blanchard01a}
from a purely classical point of view.
They found that transverse heating, Eq.~(\ref{eqn:transheat}),
of an ensemble of classical particles on average \textit{always} occurs as
they propagate through a bend, and derived
two formulas for the average transverse heating induced by a circular bend
(shown in S.I. units for later comparison):
\begin{align}
\label{eqn:bz1}
\langle E^{BZ1}_h \rangle &= \frac{2 m v^4_z}{\omega^2 \rho_0^2}
                \sin^2\Big(\frac{\omega \rho_0\phi_0}{v_z}\Big) \; , \\
\label{eqn:bz2}
\langle E^{BZ2}_h \rangle &= \frac{m v_z^4}{\omega^2 \rho_0^2} \; .
\end{align}
Here, $\langle E^{BZ1}_h \rangle$ assumes small longitudinal velocity spread
relative to $v_z$, while $\langle E^{BZ2}_h \rangle$ takes into account the
washing out of $\langle E^{BZ1}_h \rangle$ due to a significant spread in $v_z$.

\begin{figure}[th]
\includegraphics[width=3.5in]{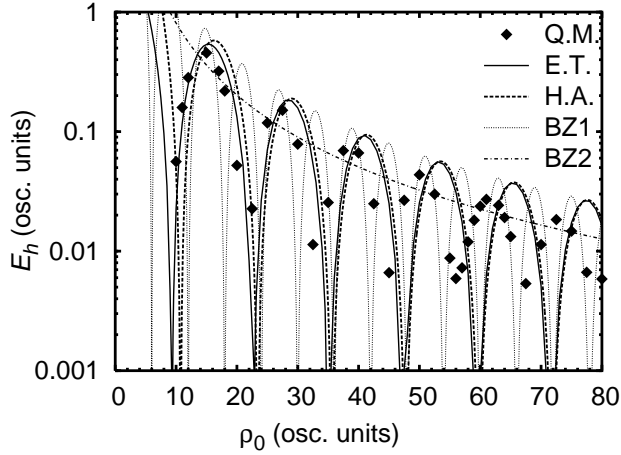}
\caption[]{ \label{fig:energyclassyrho}
Transverse heating (in oscillator units) for a $90^\circ$ bend
with fixed incoming $v_z=3$ as a function of $\rho_0$.  The diamonds
are from the time-dependent wavepacket calculations, while the
Ehrenfest trajectory results form the solid line.
The dashed line corresponds to an analytic result using a harmonic
approximation $E^{H.A.}_h$.  The dotted and dot-dashed
lines correspond to the analytic results of \cite{blanchard01a},
$\langle E^{BZ1}_h \rangle$ and $\langle E^{BZ2}_h \rangle$, respectively.
}
\end{figure}

The amount of transverse heating of a quantum wavepacket for
a series of $90^\circ$ bends with various $\rho_0$ and fixed initial
velocity $v_z=3$ are shown as the diamonds in
Fig.~\ref{fig:energyclassyrho}.   The transverse heating, Eq.~(\ref{eqn:transheat}),
for the Ehrenfest trajectory is shown as the solid line,
while three analytic results are also given: $E^{H.A.}_h$ (derived below),
$\langle E^{BZ1}_h \rangle$ and $\langle E^{BZ2}_h \rangle$.
The calculations shown in Figs.~\ref{fig:projcompilev03000} and \ref{fig:bendcircposy}
are included, along with additional calculations to bring out the dependence on $\rho_0$.
There is extremely good agreement between the wavepacket and the corresponding
Ehrenfest trajectory at small $\rho_0$, while the present results disagree
with both of the BZ formulas over the entire range of $\rho_0$.

Quantum mechanically, the transverse heating of a wavepacket, \textit{i.e.} mode-transfer,
is due to the matching between the lead and bend \cite{bromley03b} and
is manifested as the excitations seen in Fig.~\ref{fig:projcompilev03000}.
The excitation and interference of the various modes was studied
in \cite{bromley03b} as a function of $\rho_0,\phi_0$ and $v_z$, and
is not examined here in any detail.  Despite the complexity in
the quantum heating mechanism, there is close agreement between the wavepacket
and Ehrenfest trajectory results for small $\rho_0$ in Fig.~\ref{fig:energyclassyrho}.
The small $\rho_0$ limit is further investigated below as a function of $v_z$.

At the core, the physics of transverse heating for the Ehrenfest trajectory
can be understood in terms of the number of transverse oscillations
that occur during propagation through the bend.  Consider a harmonic
approximation to the effective bend potential of Eq.~(\ref{eqn:veff})
(this section is derived in S.I. units for generality):
\begin{equation}
\label{eqn:veffharm}
V_{eff}(\rho) \approx V(\rho_{min}) + \frac{1}{2} m \omega^2_{eff}(\rho-\rho_{min})^2 \; ,
\end{equation}
where $\rho_{min} \approx \rho_0 (1+\alpha^2)$ and $\omega_{eff} \approx \omega(1+\frac{3}{2}\alpha^2)$,
given $\alpha=\ell_y(m\omega\rho_0^2)^{-1}$ in our approximation
is small ($\alpha \ll 1$).
Within the bend, the Ehrenfest trajectory thus follows
\begin{equation}
\label{eqn:rhoharm}
\frac{\rho(t)}{\rho_0} = 1+2\alpha^2 \sin^2\Big[\big(1+\frac{3}{2} \alpha^2\big) \frac{\omega t}{2}\Big] \; .
\end{equation}
To determine the time, $\tau$, that the particle exits the bend at $\phi(\tau)=\phi_0$
one integrates $d\phi/dt = \ell_y/(m\rho^2(t))$ using $\ell_y=m v_z \rho_0$,
yielding $\tau \approx \rho_0 \phi_0/v_z$.
As the Ehrenfest trajectory always enters the bend with zero transverse energy,
the transverse heating is determined as the transverse energy at $t=\tau$:
\begin{equation}
\label{eqn:harmonious}
\begin{split}
E_h^{H.A.} &= \frac{1}{2} m \dot{\rho}^2(\tau) + \frac{1}{2} m \omega^2 ( \rho(\tau)-\rho_0 )^2 \\
&\approx \frac{2m v_z^4}{\omega^2 \rho_0^2} \sin^2\Big[
   \Big(1+\frac{3}{2}\big(\frac{v_z}{\omega \rho_0}\big)^2\Big) \frac{\omega \rho_0 \phi_0}{2v_z} \Big] \; .
\end{split}
\end{equation}
The agreement of $E_h^{H.A.}$ with the Runga-Kutta solution of the classical equations
of motion is seen in Fig.~\ref{fig:energyclassyrho} to be extremely good, although a
discrepancy creeps in for small $\rho_0$ where the asymptotic expansion, which
assumes $\alpha \ll 1$, breaks down.

The positions of the transverse heating maxima and minima for the Ehrenfest trajectory
are determined by $\sin^2(\rho_0 \phi_0/2v_z)$ in $E_h^{H.A.}$.
When the Ehrenfest trajectory exits the bend at the outer turning point
of $V_{eff}$ the heating is a maximum \textit{i.e.} when the
parameters $\rho_0 \phi_0/(2v_z) \approx (n+1/2)\pi$.
Exiting at the outer turning point ensures the maximum amplitude of the transverse
oscillation in the exit leads (\textit{c.f.} Fig.~\ref{fig:bendcircposy}).
For the present $90^\circ$ bends with fixed $v_z=3$, the $E_h^{H.A.}$ maxima
are $\rho_0=12$ apart, starting at $\rho_0=6$.
The transverse heating is minimized for $\rho_0 \phi_0/(2v_z) \approx n\pi$,
where the trajectory exits the bend at $\rho(\tau)=\rho_0$. Of course,
at $x=\rho_0$ in the exit lead there is no force acting on the
Ehrenfest trajectory, and thus no transverse heating.
The $E_h^{H.A.}$ transverse heating minima in Fig.~\ref{fig:energyclassyrho} occur every
multiple of $\rho_0=12$ for large $\rho_0$, with a slight departure for the
two minima located near $\rho_0 \approx 23$ and $10$ due to the corrective term
in Eq.~(\ref{eqn:harmonious}).

There are two complicating factors hidden amongst the
quantum and Ehrenfest comparisons: dispersion and mode-excitation. 
Dispersion is manifested in the wavepacket calculations as a dampening
of the oscillations of $E_h$ with $\rho_0$ compared to the
Ehrenfest trajectory results.
Mode-excitation also complicates the comparison since each mode that is
excited upon entering the bend executes a different number of transverse
oscillations through the bend.  For $v_z=3$, both of these velocity effects
are not so significant for small $\rho_0$, but begin to play a role at larger $\rho_0$.
While Ehrenfest's theorem is exact, the present results demonstrate that
the approximation of Eq.~(\ref{eqn:ehren}) begins to breakdown in the
large $\rho_0$ regime, where wave effects such as dispersion become significant.

The variance of the BZ results with the Ehrenfest trajectory calculations
is seen in comparing Eq.~(\ref{eqn:harmonious}) with Eq.~(\ref{eqn:bz1}).
This clearly shows a factor of $2$ difference in the sinusoidal
periodicity (apart from an additional corrective term),
with the $\langle E^{BZ1}_h \rangle$ formula predicting transverse heating
minima at every multiple of $\rho_0 = 6$ (for $90^\circ$ bends with fixed $v_z=3$).
To understand the difference in the periodicity, consider a single
classical particle entering at the center of the bend
with \textit{non-zero} transverse velocity: this particle
will pass through $\rho=\rho_0$ twice as it oscillates
through one period of the effective potential, Eq.~(\ref{eqn:veff}).
If the exit point out of the bend occurs at $\rho(\tau)=\rho_0$,
the particle exits with the same transverse speed that it entered with.
This means that no transverse heating of the classical particle occurs
for two points during one oscillation in the effective bend potential.
In the gentle bend limit, the heating minima would indeed occur
every $\rho_0 = 6$.
On the other hand, the Ehrenfest trajectory, which strictly maintains
$\rho(t) \ge \rho_0$ throughout the bend, requires one complete
oscillation in the bend potential to reach back to $\rho(t) = \rho_0$,
(and thus heating minima occur every $\rho_0=12$ in the large $\rho_0$ limit).

The comparison of the BZ results with the quantum wavepacket calculations
seen in Fig.~\ref{fig:energyclassyrho} also warrants further comment.
The transverse heating minima for quantum wavepackets drift away
from the Ehrenfest trajectory minima as $\rho_0$ becomes large
in Fig.~\ref{fig:energyclassyrho}.  In this limit, it might be
expected that the ''gentle bend'' assumption leading
to Eq.~(\ref{eqn:bz1}) is better satisfied.
We do not, however, observe the appearance of two minima in the
quantum wavepacket calculations as per $\langle E^{BZ1}_h \rangle$.
The BZ formulae are based on an analysis of an ensemble
of classical particles that enter relatively gentle bends
with transverse velocities much smaller than the propagation velocity.
To examine the gentle bend, high velocity wavepacket limit is not only
computationally taxing, but remains beyond the scope of the present paper.
While the present quantum results do not agree with either of Blanchard and
Zozulya's formulae, it must be emphasized that the magnitude
of the five different results shown in Fig.~\ref{fig:energyclassyrho}
all follow the same decay trend: $E_h \propto \rho_0^{-2}$.

\begin{figure}[th]
\includegraphics[width=3.5in]{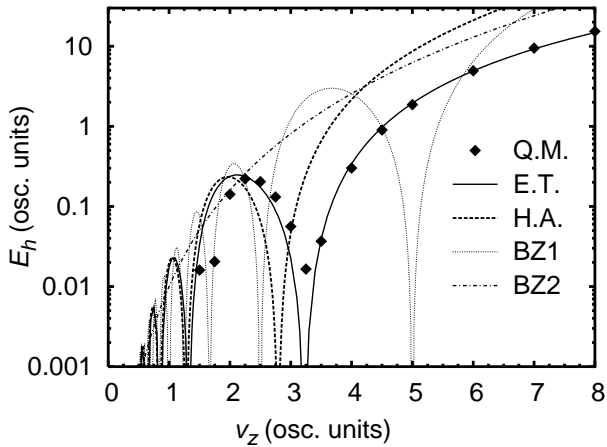}
\caption[]{ \label{fig:energyclassyv}
Transverse heating (in oscillator units) for a $90^\circ$ bend
with fixed $\rho_0=10$ as a function of the incoming propagation velocity.
The diamonds are from the time-dependent wavepacket calculations,
while the Ehrenfest trajectory results form the solid line.
The dashed line corresponds to an analytic result using a harmonic
approximation $E^{H.A.}_h$.  The dotted and dot-dashed lines
correspond to the BZ analytic results,
$\langle E^{BZ1}_h \rangle$ and $\langle E^{BZ2}_h \rangle$, respectively.
}
\end{figure}

Finally, the dependence on velocity for a tight ($\rho_0=10$) bend can be seen
in Fig.~\ref{fig:energyclassyv}.
There is striking correspondence between the transverse heating of
the quantum wavepackets and Ehrenfest trajectories for $v_z>2$,
even up to the highest energy, $v_z=8$, where the quantum calculation
has 32 energetically open modes.  This agreement follows one
of the fundamental postulates of quantum mechanics, Bohr's correspondence
principle, which states that in the limit of high quantum number, the quantum
and the classical pictures merge.  This is well known for a 1-D SHO,
and it is nice to see that it applies to the present potentials.
The correspondence is completely lacking, of course, at low-energies since
only small quantum numbers are allowed.  In particular, no
transverse heating of the wavepacket is allowed for $v_z\le\sqrt{2}$ since only
the ground state mode is accessible.

On the basis of the harmonic approximation in Eq.~(\ref{eqn:harmonious}), the
$E_h$ minima for $\rho_0=10$ occur at $v_z=2.5,1.25$, etc.  The Ehrenfest
trajectory calculations deviate from this periodicity as $v_z$ increases, since
it is a tight bend in which there is a significant perturbation
of the effective bend potential of Eq.~(\ref{eqn:veff}) due to the
centripetal term.  The BZ formula Eq.~(\ref{eqn:bz1}) predicts additional minima
at $v_z=5,5/3$, etc, and Fig.~\ref{fig:energyclassyv} shows a complete departure
of the BZ predictions from the quantum and Ehrenfest results across the entire
range of $v_z$.  This is not unexpected since it was already seen in the
$v_z=3$, tight $\rho_0$ limit in Fig.~\ref{fig:energyclassyrho}.

\section{Conclusions}

The time-dependent Schr{\"o}dinger equation has been solved for
non-interacting, low-energy wavepackets propagating through circular bends.
Our main goal for the present time-dependent calculations was to
understand the conditions under which the classical picture of
atom propagation through microstructures is valid for predicting transverse
excitation.  To this end, a series of time-dependent quantum
and Ehrenfest trajectory calculations were performed and compared.

In the tight-bend limit, $\rho_0 \le 30$, for $v_z=3$ there was good
agreement between the transverse heating predicted by the wavepacket and the
Ehrenfest trajectory calculations.  In contrast to the purely
SHO case, the path of the wavepacket averages
$\langle x \rangle(t)$,$\langle z \rangle(t)$ through the bend
were found to be completely different from the Ehrenfest trajectory
for small $\rho_0$.  The peak probability density of the wavepacket, however,
wriggles out of the bend along the same path as the Ehrenfest trajectory.
Bohr's correspondence principle is demonstrated in the tight-bend,
high propagation velocity limit (where high-$n$ modes can be excited)
where great agreement is seen between the transverse heating
of the wavepacket and its Ehrenfest trajectory.

The agreement for larger $\rho_0$ was not as good due to the
increasingly important effects of wavepacket dispersion.
In general, as long as the propagation velocity is such that there
are minimal wave effects (such as reflection and dispersion) the
approximation to Ehrenfest's theorem introduced in
Eq.~(\ref{eqn:ehren}) provides a useful approximation to
the exact Ehrenfest's theorem and thus the quantum mechanics.

A note of caution should be added when extending this idea to
other microstructures.  For example, a wavepacket exiting
an abruptly terminating transverse potential can transfer
transverse energy into longitudinal energy, leading to
transverse cooling \cite{jaaskelainen02c}.
In that particular study, an ensemble of classical particles
was found to reproduce the wavepacket expectation values
(given situations where quantum reflections did not play a role).
The Ehrenfest trajectory through such a potential, however, would be
completely unaffected by the changes in the transverse potentials.

As the temperature of the atoms used in atom chip experiments
decreases, the modelling of experiments using Monte-Carlo simulations 
will increasingly need to take into account the wave nature
of the atoms.  The present calculations provide guidance as to when
this can simply be achieved by using a single classical trajectory
to calculate the quantum observables of waves propagating through
simple microstructures.

\begin{acknowledgments}
This work was supported by the Department of the
Navy, Office of Naval Research, and by the Research Corporation.
\end{acknowledgments}

\appendix

\section{Finite differencing}
\label{sec:finite}

Most implementations of finite differences for the Schr\"odinger
equation implicitly require a volume element that is constant over the whole space.  
This condition is trivially satisfied for Cartesian coordinates and
can usually be achieved through a rescaling of the wave function for
other coordinate systems.  Spherical coordinates are a prime example
of the latter case since the radial wave function is often scaled
by a factor of $r$ to remove the first derivative from the kinetic 
energy, effectively putting the system into a Cartesian coordinate
system with a boundary condition.

Another common coordinate system, cylindrical coordinates, is not so easily
handled --- rescaling the wave function involves a factor of $\sqrt{\rho}$.
The overall behavior of the wave function near $\rho$=0 is then non-analytic,
making finite differencing invalid since differencing relies on Taylor series
expansions.  Further, straightforward differencing of the unscaled $\rho$ equation
gives a non-Hermitian operator, which is a situation to be avoided in general.
The angular momentum operator in spherical coordinates suffers from
similar problems upon rescaling.  Somewhat {\it ad hoc} schemes have been
formulated to deal with these problems, but are less than satisfactory
from the viewpoint of wanting a general differencing scheme applicable to
arbitrary coordinate systems.

One solution to this problem has actually been known for quite some time,
and is based on general principles of discrete calculus and differencing \cite{mitchell80a}.
Our approach is similar to Ref.~\cite{witthoeft03a}, however, and applies the
more familiar variational principle to derive the differencing equations.
The derivation will be outlined for the one-dimensional time-independent equation for
simplicity, but carries through in exactly the same way for the
multi-dimensional time-dependent equation.  We begin with the energy functional
\begin{equation}
\label{EFunc}
E=\frac{\int \left( \frac{\hbar^2}{2\mu} \frac{d \psi^*}{dx} \frac{d \psi}{dx}+V\psi^*\psi\right) \rho(x) dx}
{\int \psi^* \psi \rho(x) dx} \; .
\end{equation}
Note that in writing the kinetic energy as we have will guarantee that our
resulting difference representation is Hermitian.  To proceed, we must
choose quadrature and differencing rules for the integrals and derivatives,
respectively.  We choose the midpoint rule
\begin{equation}
\label{IntRule}
\int f(x) dx \longrightarrow \sum_{i=1}^N f(x_i) (x_{i+\frac{1}{2}}-x_{i-\frac{1}{2}}) \; ,
\end{equation}
and central differencing
\begin{equation}
\label{DerivRule}
\left. \frac{d f}{dx}\right|_{x_i} \longrightarrow
\frac{f_{i+\frac{1}{2}}-f_{i-\frac{1}{2}}}{x_{i+\frac{1}{2}}-x_{i-\frac{1}{2}}} \; .
\end{equation}
The grid points have been left arbitrary to allow maximum flexibility of
the representation.  A uniform grid can, of course, be chosen.

Substituting Eqs.~(\ref{IntRule}) and (\ref{DerivRule}) into Eq.~(\ref{EFunc}) and
minimizing with respect to the wave function on the grid,
\begin{equation}
\frac{\partial E}{\partial \psi_j^*}=0 \; ,
\end{equation}
gives the difference equation
\begin{equation}
\begin{split}
\frac{\hbar^2}{2\mu}&\left[ \frac{\psi_j\!\!-\!\!\psi_{j-1}}{x_j\!\!-\!\!x_{j-1}}\rho_{j-\frac{1}{2}}\!\!-\!\!
                     \frac{\psi_{j+1}\!\!-\!\!\psi_j}{x_{j+1}\!\!-\!\!x_j}\rho_{j+\frac{1}{2}} \right] \\
&+V_j \psi_j \rho_j (x_{j+\frac{1}{2}}\!\!-\!\!x_{j-\frac{1}{2}})
= E \psi_j \rho_j (x_{j+\frac{1}{2}}\!\!-\!\!x_{j-\frac{1}{2}}) \;,
\end{split}
\end{equation}
for $j=1,\ldots,N$. Writing this as a matrix equation,
\begin{equation}
\label{GenMatrixEqn}
{\bf H} \vec{\psi} = E {\bf S} \vec{\psi} \; ,
\end{equation}
shows that our differencing scheme does indeed yield Hermitian operators.  The
Hamiltonian matrix is tridiagonal just as for the usual second-order finite differencing
of the Schr\"odinger equation; the overlap matrix $\bf S$ is diagonal.  

This difference scheme must be supplemented by boundary conditions.  If the wave
function is to be zero on the boundary, then $\psi_0$ (or $\psi_{N+1}$) must be
set to zero.  If the derivative of the wave function is to be zero, then two
grid points should be chosen to straddle the boundary point and the condition
$\psi_0$=$\psi_1$ imposed (or $\psi_N$=$\psi_{N+1}$).  This condition requires that
\begin{equation}
\frac{\hbar^2}{2\mu} \frac{\rho_\frac{1}{2}}{x_1-x_0}
\end{equation}
be subtracted from the first diagonal element of $\bf H$; a similar term should
be subtracted from the last element to apply the boundary condition at the other
boundary.  The case in which no boundary condition is required --- if $\rho$=0 ---
is treated the same as the zero derivative boundary condition.  In all cases, the 
values of $\rho_i$ outside of the grid should be chosen to be symmetric with respect
to the boundary.

This equation can be put in somewhat more convenient form using the usual transformation
for generalized eigenvalue problems,
\begin{equation}
{\bf S} = {\bf L}^T {\bf L} \;,
\end{equation}
which is always possible for positive definite matrices $\bf S$.  In this case, the result
is trivial since $\bf S$ is diagonal:
\begin{equation}
L_{i} = \sqrt{S_i} \; .
\end{equation}
Equation~(\ref{GenMatrixEqn}) can then be transformed with the relations
\begin{equation}
\tilde{\bf H} = {\bf L}^{-1} {\bf H} {\bf L}^{-T} \quad \mathrm{and} \quad \vec{\phi} = {\bf L}^T \vec{\psi}
\end{equation}
into the standard eigenvalue problem
\begin{equation}
\tilde{\bf H} \vec{\phi} = E \vec{\phi} \; .
\end{equation}
Other transformations will produce a standard eigenvalue problem from the
generalized one, but this transformation was chosen since it produces a
Hermitian Hamiltonian.  Explicitly, the nonzero matrix elements of $\tilde{\bf H}$ are
\begin{equation}
\tilde{H}_{ii} = \frac{\hbar^2}{2\mu} \left[\frac{\rho_{i-\frac{1}{2}}}{x_i-x_{i-1}}+\frac{\rho_{i+\frac{1}{2}}}{x_{i+1}-x_i}
\right] \frac{1}{\rho_i (x_{i+\frac{1}{2}}-x_{i-\frac{1}{2}})}+V_i
\end{equation}
and
\begin{equation}
\begin{split}
\tilde{H}_{i,i+1} = -\frac{\hbar^2}{2\mu} &\frac{\rho_{i+\frac{1}{2}}}{x_{i+1}-x_i} \times \\
  &\frac{1}{\sqrt{\rho_i (x_{i+\frac{1}{2}}-x_{i-\frac{1}{2}}) \rho_{i+1} (x_{i+\frac{3}{2}}-x_{i+\frac{1}{2}})}}.
\end{split}
\end{equation}
The elements $\tilde{H}_{i,i-1}$ can be obtained from the symmetry of $\tilde{\bf H}$.

With our quadrature rule, normalization takes the form
\begin{equation}
\sum_{i=1}^N |\psi_i|^2 \rho_i (x_{i+\frac{1}{2}}-x_{i-\frac{1}{2}}) = 1 \; .
\end{equation}
Consequently, the transformed wave function satisfies 
\begin{equation}
\sum_{i=1}^N |\phi_i|^2  = 1 \; ,
\end{equation}
since the transformation includes both the volume element and step size.  Care
must be taken to use the physical wave function $\vec{\psi}$ for
calculations or, alternatively, to derive the equivalent expressions for the transformed
function $\vec{\phi}$.

The transformed Hamiltonian $\tilde{\bf H}$ and wave function $\vec{\phi}$ can be 
inserted directly into the Crank-Nicolson time propagation scheme in place of
the usual uniform Cartesian difference Hamiltonian.

\section{Expectation values for SHO superposition states}
\label{sec:timedeps}

The time-dependence of the expectation values of time-independent operators $\Theta$
can be determined in quantum mechanics from the Heisenberg equation of motion:
\begin{equation}
\label{eqn:timecommut}
\langle \Theta \rangle(t) = \frac{1}{i\hbar} \langle[\Theta,H]\rangle \; .
\end{equation}
For energy eigenstates $\psi_E$, $\langle\psi_E |[\Theta,H]| \psi_E\rangle=0$,
and thus $\langle \Theta \rangle$ is time-independent.  For a superposition
of such states, both the time and spatial dependence of the expectation values is
not so simple and is worth this brief appendix.

In our analysis, the transverse heating played the critical role in our comparisons
between the quantal and classical results.  The transverse heating is simply the
energy transferred from the wavepackets longitudinal kinetic energy into
transverse energy.  Quantum mechanically, of course, we must take the expectation
value of the kinetic energy operator.  One might be tempted to use the virial theorem
to more simply calculate this using twice the expectation value of the potential energy,
however, it is well known \cite{merzbacher70a} that the virial theorem only
strictly holds for stationary states.

It turns out that for a simple harmonic oscillator in one dimension that
the expectation value of the kinetic energy for an arbitrary state is
\begin{align}
\langle \Psi | T | \Psi \rangle(t) &= \frac{1}{2} \sum_{n=0}^\infty \Big[ (n+\frac{1}{2}) |b_n|^2 \nonumber\\
             &-\sqrt{(n+2)(n+1)} \: \mathrm{Re} \big(e^{2 it } b^*_{n+2} b_n\big)\Big] \; 
\end{align}
where $b_n$ are the expansion coefficients of $|\Psi\rangle$ on the SHO
eigenstates $|n\rangle$.  The expectation value of kinetic energy thus oscillates
about half the average total energy.  Upon time-averaging, the second term falls out,
so that the virial theorem holds on average.  Since only the time-average will
usually be important, this result can be handy for numerical calculations.
Note that if only states with opposite
parity or indices separated by more than two are populated, then the virial
theorem holds for all times.

For waveguide applications such as those considered in this paper,
it is convenient to write the wave function as
\begin{equation}
\Psi(x,z,t) = \sum_{n=0}^\infty \psi_n(z,t)  \varphi_n(x) e^{-i E_n t}
\end{equation}
where $\varphi_n(x)$ is the transverse SHO wave function.
This multi-moded wavepacket can then be used to calculate observables.
For instance, the transverse kinetic energy is
\begin{align}
\label{eqn:TransPos_of_x}
\langle \Psi | T_x | \Psi \rangle(t) &= \frac{1}{2} \sum_{n=0}^\infty \Big[ (n+\frac{1}{2}) \int \!\! dz\,|\psi_n(z,t)|^2 \nonumber\\
             &-\sqrt{(n+2)(n+1)} \: \mathrm{Re} \big(e^{2 it }\!\! \int \!\! dz \, \psi_{n+2}^*(z,t) \psi_n(z,t)\big)\Big] \; ,
\end{align}
and the transverse position is
\begin{gather}
\label{eqn:TransPos_of_t}
\langle \Psi | x | \Psi \rangle(t) = \sqrt{2}\sum_{n=0}^\infty
     \sqrt{n+1} \: \mathrm{Re} \big(e^{i t}\!\!  \int\!\! dz \,
         \psi^*_{n+1}(z,t) \psi_n(z,t) \big) \; .
\end{gather}
So, as long as the components $\psi_n$ overlap, they can interfere, leading to
time-varying observables.  As soon as the components no longer overlap --- and there
is no mechanism for mode conversion --- the observables will become time-independent
as expected.

We can also use these ideas to interpret the wave function at a fixed time by 
only integrating over $x$.  The transverse position at a given time $t$
is then
\begin{gather}
\label{eqn:TransPos_of_z}
\langle \Psi | x | \Psi \rangle(z,t) = \sqrt{2}\sum_{n=0}^\infty
     \sqrt{n+1} \: \mathrm{Re} \big(e^{i t}  \, \psi^*_{n+1}(z,t) \psi_n(z,t) \big) \; 
\end{gather}
as a function of $z$.  This expression can be further simplified by considering
the plane-wave limit in which each component, $\psi_n$, has large longitudinal extent
(and small spread in momentum) since we can then replace
$\psi_n$ by $b_n e^{ik_n z} e^{-i\frac{1}{2}k_n^2 t}$:  
\begin{equation}
\langle \Psi | x | \Psi \rangle(z,t) = \sqrt{2}\sum_{n=0}^\infty
     \sqrt{n+1} \: \mathrm{Re} \big(b^*_{n+1} b_n\, e^{i (k_n-k_{n+1}) z} \big) \; 
\end{equation}
where $k_n=\sqrt{2 (E_T-n)-1}$ and $E_T$ is the total energy
(which is the same for each mode).  This time-independent expression can be used, for instance, to
understand the wriggles seen in Fig.~\ref{fig:projcompilev03000} as the
wavepacket exits the circular bend.   Where two modes overlap, the position of
the resulting $\langle \Psi | x | \Psi \rangle(z)$ wriggles will be independent of time
as the wavepacket propagates (like a snake threading its body through
a single curved path).

\begin{figure}[thp]
\includegraphics[width=3.5in]{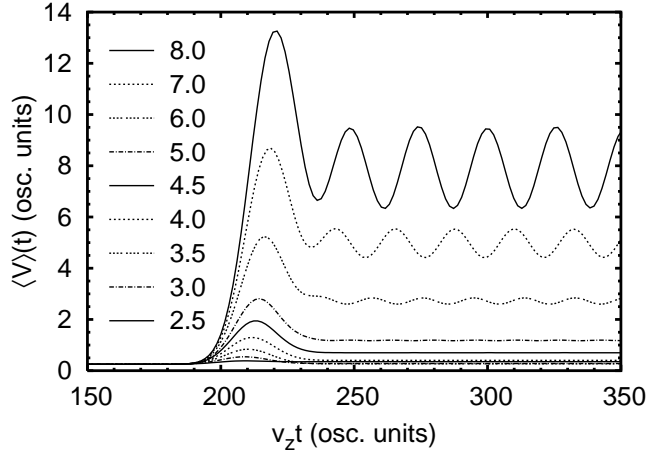}
\caption[]{ \label{fig:potentialtime}
$\langle V \rangle$ (in oscillator units) as a function of time
for a $90^\circ$ bend with fixed $\rho_0=10$ for incident wavepackets
with varying initial velocity, $v_z$, ranging from $2.5$ to $8.0$.  Each wavepacket is
initially in the ground state of the SHO ($\langle V \rangle(t=0) \approx 0.25$), and
transverse heating by the bend is seen as an increase in $\langle V \rangle$.
The time axis has been rescaled by $v_z$ so that the effects due to the
bend (seen as the peak in $\langle V \rangle(t)$) occurs at roughly
the same point in the figure. 
}
\end{figure}

Fig.~\ref{fig:potentialtime} shows the average potential
energy of wavepackets initially in the ground state of the SHO
with varying incident velocity $v_z$
as they propagate through a $90^\circ$ circular bend of radius $\rho_0=10$.
As the wavepacket is sloshing around in the bend, the potential energy increases
and settles into an oscillation between kinetic and potential
energies in the exit lead.  As $t\to\infty$, the components $\psi_n$ of
the wavepacket will no longer overlap [see Eqs.~(\ref{eqn:TransPos_of_x})
and (\ref{eqn:TransPos_of_t})].
Thus, the expectation values will no longer oscillate, and the
oscillations seen in Fig.~\ref{fig:potentialtime} will dampen with time
towards half of the total transverse energy.  For the calculations seen here,
this would involve propagating the wavepackets to stupidly large distances down
the exit lead.  Instead, to determine the transverse heating, we used averages
of $\langle T \rangle$ and $\langle V \rangle$ over one oscillation once the
wavepacket has completely exited the bend,
\textit{i.e.} the kinetic energy in the propagation direction is given by
$\langle T_z \rangle = \langle T \rangle_{avg} - \langle V \rangle_{avg}$,
and the transverse heating is
$E_h = \langle T_z \rangle(t \to \infty) - \langle T_z \rangle(t \to -\infty)$.

The present Crank-Nicolson with finite difference calculations
also has tiny oscillations between $\langle V \rangle$ and $\langle T \rangle$,
while the total energy of the wavefunction is being conserved
(typically to better $10^{-6}$ osc. units).  
These oscillations even occur during propagation through a straight
SHO-based waveguide.   This is due to using an initial wavepacket defined
by Eq.~(\ref{eqn:gaussinit}) with $\Delta_x=1$ centered at $x_0=\rho_0$, which is
not the exact eigenstate of the finite difference representation.   These
numerical oscillations are kept to a minimum (here less than $10^{-3}$ osc. units,
and not visible on Fig.~\ref{fig:potentialtime}) using a
dense, non-equally spaced transverse grid.

\end{document}